    \documentclass[oneside,12pt]{article}

    \title{Stochastic quantisation of locally supersymmetric models}

    \author{Hossein Farajollahi \& Hugh Luckock\\ {\small School of
    Mathematics and Statistics}\\
    {\small University of Sydney}\\
    {\small NSW 2006, Australia} }

    \date{\today}

\begin{document}
    \maketitle

    \begin{abstract}

    Stochastic quantisation normally involves the introduction of a
    fictitious extra time parameter, which is taken to infinity so that the
    system evolves to an equilibrium state.

    In the case of a locally supersymmetric theory, an interesting new possibility
    arises due to the
    existence of a Nicolai map. In this case it turns out that no additional time
    parameter is required, as the existence  of the Nicolai map ensures that the
    same job can be done by the existing time parameter after Euclideanisation.
    This provides the quantum theory with a natural probabilistic
    interpretation, without any reference to the concept of an inner product or a Hilbert space structure.

    \end{abstract}

    \newpage

    \def\be{\begin{equation}}
    \def\ee{\end{equation}}
    \def\bea{\begin{eqnarray}}
    \def\eea{\end{eqnarray}}
    \def\M{{{\cal M}}}
    \def\bdy{{\partial\cal M}}
    \def\w{\widehat}
    \def\n{\widetilde}
    \def\real{{\bf R}}
    \def\q{{\bf q}}

    \section{Introduction}

    In 1981 Parisi and Wu proposed a radical new way to obtain Green's
    functions in Euclidean quantum field theory \cite{Parisi:1}.
    Known as stochastic quantisation, this approach enjoys certain
    conceptual and practical advantages over more traditional ones.
    In this paper, we show how this approach arises naturally
    for locally supersymmetric models and provides them with a
    natural probabilistic interpretation of the wave function
    without reference to the concept of an inner product.

    Stochastic quantisation normally requires the introduction of
    an extra parameter $t$, referred to as the fictitious time. Evolution
    with respect to $t$ is assumed to be stochastic, governed
    by a type of Langevin equation which gradually forces the system
    into thermal equilibrium. The static probability distribution
    arising in the $t\to\infty$ limit can then be identified with
    the Euclidean path integral measure and used to calculate the
    Euclidean Green functions.

    In the case of a supersymmetric theory, an interesting new
    possibility arises due to the existence of a Nicolai map. (This is
    a transformation which converts any supersymmetric
    theory into a non-interacting bosonic one \cite{Nicolai}, and which
    generally has the form of a stochastic differential equation in Euclidean theories.)
    In this case, there is no need to introduce a fictitious time parameter;
    the presence of the Nicolia map means that the same role can be performed
    quite satisfactorily by the physical time parameter from the original
    theory.

    Regarding the Nicolai map as a stochastic process leads to an
    interpretation of certain components of the wave function
    (in an appropriate representation) as probability densities whose integral
    is conserved in Euclidean time. This provides a simple stochastic
    interpretation for supersymmetric quantum theories, without the need
    to define an inner product on the space of states or appealing to the
    concept of a Hilbert space. This is particularly useful in the context
    of quantum cosmology, where it is difficult to identify the inner
    product \cite{Graham:1}.

    In general, finding Nicolai maps for supersymmetric theories is not an
    easy task. In fact, it may be easier to find the associated Fokker-Planck
    equation directly, thus by-passing the difficulties of explicitly
    constructing Nicolai maps.

    This paper is organised as follows: A review of stochastic
    quantisation of a scalar field is given in \S \ref{StochQuant}. For a
    supersymmetric theory, the notion of Nicolai map is introduced in \S\ref{Nicolai}.
    In \S\ref{Canonical}, the canonical formulation of a one-dimensional supersymmetric non-linear
    $\sigma$ model is  developed. Supersymmetry is always broken
    when boundaries are present.
    In \S\ref{BoundaryTerms} we will see that there is a boundary correction to
    the action that restores invariance under a sub-algebra of
    supersymmetry transformations. Derivation of Nicolai map and Fokker-Planck equation
    in case of a rigid supersymmetry is done in
    \S\ref{FokkerPlanck}. In this section, it also has been discussed that the
    result can be obtained directly from canonical quantisation.
    Section \S\ref{localSUSY} is devoted to quantisation of a
    locally supersymmetric model.

    \section{Stochastic quantisation of scalar field theories}\label{StochQuant}

    The main idea in stochastic quantisation is to view Euclidean quantum field
    theory as the
    equilibrium limit of a statistical system coupled to a thermal reservoir. This
    system is assumed to evolve with respect to a fictitious time variable $t$
    and approaches the
    equilibrium limit as $t\rightarrow \infty$. The coupling to a heat
    reservoir is simulated by means of a stochastic noise field which causes
    the original Euclidean field to wander randomly on its manifold. In the
    equilibrium limit stochastic averages become identical to ordinary Euclidean
    vacuum expectation values.

    In this section we outline the stochastic,
    quantisation of scalar field theory using the approach of Parisi and
    Wu\cite{Parisi:1}. For a more complete discussion, the reader is referred to the
    review article by Damgaard and Huffel \cite{Damgaard:1}.

    The basic idea is as follows:

    \begin{itemize}
    \item [{\bf i})] We imagine that each field $\phi (x)$ depends on an additional
    coordinate, the
    fictitious time t
    \be
    \phi (x)\rightarrow \phi (x,t).
    \ee

    \item [{\bf ii})] We suppose that the evolution of $\phi (x, t)$ with respect
    to the fictitious time $t$ is described
    by a stochastic differential equation that allows for relaxation to
    equilibrium. Specifically, one postulates that the evolution is governed by
    the Langevin equation
    \be
    \frac{\partial \phi (x,t)}{\partial t}=-\frac{\delta S}{\delta \phi (x)}+\eta
    (x,t),\label{eq:Lange equation}
    \ee
    where $S[\phi]$ is the Euclidean action and $\eta (x,t)$ is a Gaussian
    white noise with correlations
    given by

    \bea
    &&\langle\eta(x,t)\rangle_\eta =0\nonumber\\
    &&\langle\eta(x_1,t_1)\eta(x_2,t_2)\rangle_\eta=2\delta^n
    (x_1-x_2)\delta(t_1-t_2)
    \eea

    \item [{\bf iii})] Given some initial condition at $t=t_0$ and some realisation
    of
    $\eta(t)$,equation (\ref{eq:Lange equation}) has a unique solution $\phi_\eta (x,t)$. The
    correlation functions of $\phi_\eta$ are then obtained as Gaussian
    averages over all possible realisations of $\eta(t)$;
     \bea
    \langle\phi_\eta (x_1,t_1)\cdots \phi_\eta (x_k,t_k)\rangle_\eta =\frac{\int
    D\eta e^{-\frac{1}{4}\int d^n x dt \eta^2(x,t)}\phi_\eta (x_1,t_1)\cdots
     \phi_\eta (x_k,t_k)}{\int D\eta e^{-\frac{1}{4}\int d^n x dt \eta^2(x,t)}}
    \eea
    As $t\rightarrow 0$, equilibrium is reached, and the (equal
    time) correlation functions of $\phi_\eta$ tend to the corresponding
    quantum green functions
    \be
    \lim_{t\rightarrow \infty}\langle\phi(x_1,t)\cdots \phi(x_k,t)\rangle_\eta
    =\langle\phi
    (x_1)\cdots \phi(x_k)\rangle
    \ee
    \end{itemize}
    An alternative to the Langevin approach is to study the Fokker-Planck
    formulation. In this approach, the stochastic averages are represented as
    functional integrals
    \be
    \langle\phi(x_1,t)\cdots \phi(x_k,t)\rangle_\eta =\int D\phi
    f(\phi,t)\phi(x_1)\cdots \phi(x_k)
    \ee
    where the probability density functional $f[\phi(x),t]$ is a solution to the
    Fokker-Planck equation
    \be
    \frac{\partial f}{\partial t}=\int d^n x\frac{\delta}{\delta
      \phi(x,t)}\biggl (\frac{\delta S}{\delta \phi(x,t)}+\frac{\delta}{\delta
      \phi(x,t)}\biggr )f.\label{eq:fok-plant equation}
    \ee
    The physical correlation functions are then calculated using
    the equilibrium probability density functional
    \be
    f[\phi]=\lim_{t\rightarrow \infty}f[\phi
    ,t]=f^{eq}(\phi)=\frac{e^{-S}}{\int D\phi ^{-S}}\cdot
    \ee
    If we define the operator
    \be
     O =-\int d^nx\frac{\delta}{\delta
      \phi(x,t)}\biggl (\frac{\delta S}{\delta \phi(x,t)}+\frac{\delta}{\delta
      \phi(x,t)}\biggr ),
    \ee
    then the Fokker-Planck equation (\ref{eq:fok-plant equation}) can be rewritten simply as
    \be
    \frac{\partial f (x, t)}{\partial t}=-O f(x,t).
    \ee

    One can show that the results obtained from stochastic quantisation agree
    with those obtained by more conventional methods \cite{Parisi:1}. Now for a system with a single degree of
    freedom $x$, and a potential $V(x)$. The Fokker-Planck equation
    reads
    \be
    \frac {\partial f(x,t)}{\partial t}=\frac{\partial }{\partial
    x}\biggl(\frac{\partial }{\partial
      x}+\frac{\partial V}{\partial x}\biggr )f(x,t)
    \ee

    \section{Supersymmetric models and Nicolai maps}\label{Nicolai}

    Nicolai has shown \cite{Nicolai} that there is a map which transforms any
    supersymmetric theory to a
    free bosonic theory. The simplest example of Nicolai map is the Langevin
    equation, although in
    general for a typical supersymmetric action the Nicolai map will not have
    such a simple form \cite{Graham:1}.

    The Nicolai's theorem essentially states the following: Take a supersymmetric theory and
    integrate out the fermion fields in the path integral. This contributes a
    non trivial determinant factor to the (bosonic) path integral. There now exists
    a transformation of the bosonic fields whose Jacobian determinant
    exactly cancels the determinant of the fermion integrations, and which
    simultaneously transforms the remaining bosonic part of the action to
    that of a non-interacting bosonic theory.

    The transformation (Nicolai map) will be invertible if one impose an
    appropriate number of boundary conditions on the bosonic variables in the
    theory. However, imposition of such conditions break the supersymmetry
    since the supersymmetric variation of the Lagrangian produces a total
    divergence which yields a boundary term when integrated.

    Nicolai's theorem
    will not be applicable unless the supersymmetry algebra has some
    graded sub-algebra whose bosonic generators preserve the boundary
    \cite{Luckock:2}. Note that in the framework of quantum theory, boundaries
    are important since one is generally calculating transition amplitudes
    between specified boundary data. Hence, boundary effects cannot be
    neglected.

    In general, only the Euclidean version of a supersymmetric theory
    will admit a Nicolai map which can be interpreted as stochastic
    differential equation describing the evolution of the system in Euclidean
    time. This suggests that in
    quantum theory one can interpret the wave function (in
    an appropriate representation) as a probability density function whose integral
    is
    conserved in Euclidean time. In fact, we are more interested in the static
    state which is finally
    reached, since it represents the ground state of the Euclidean theory.

    \section{N=1 supersymmetry}\label{Canonical}

    Here we present the canonical formulation of a
    one-dimensional supersymmetric non-linear $\sigma$ model
    describing a particle moving in a curved configuration space.
    This example will be used throughout the paper to illustrate
    an approach which can be applied to quite general supersymmetric
    models.

    \def\EL{L}
    Suppose the position $\q(t)$ of the particle at time $t$ is described by $n$
    coordinates $q^i(t)$. The action for the locally supersymmetric
    Euclidean model is $S=\int \EL dt$ with the Lagrangian
    \bea
    \EL & = &\frac{1}{2}\left[N^{-1}g_{ij} \dot q^i\dot q^j
    +Ng^{ij}V_{,i}V_{,j}
    +\overline\psi_i(D\psi^i) -(D\overline\psi_i) \psi^i\right]
    \nonumber\\
    &&+ NV_{;ji}\overline\psi^i\psi^j
    -\overline\chi\psi^i(g_{ij}\dot q^j +NV_{,i})-\chi\overline\psi_i
    (\dot q^i -Ng^{ij}V_{,j})\nonumber\\
    &&+ N\overline\chi\chi\overline\psi_i\psi^i
    -\frac{1}{2}NR_{ijkl}\overline\psi^i\psi^j\overline\psi^k
    \psi^l.\label{eq:basic_Lagrangian}
    \eea
    where $g_{ij}(\q)$ is the metric of the configuration space in which the
    particle moves, $V(\q)$ is a potential function defined on this space,
    and $\dot q^i\equiv dq^i/dt$.
    The covariant time derivatives of the fermion fields  are defined as
    \bea
    D\psi^i=\dot \psi^i +\Gamma^i{}_{jk}\psi^j\dot q^k\\
    D\overline{\psi}_i=\dot{\overline\psi_i}-\Gamma^j{}_{ik}
    \overline\psi_j\dot q^k
    \eea
    where $\Gamma^i{}_{jk}$ is the usual symmetric Christoffel connection on
    the configuration space, compatible with the metric $g_{ij}$ \cite{Mukhi}.
    The Riemann curvature tensor is
    \bea
    R^i{}_{jkl}=
    \partial_k\Gamma^i{}_{jl}
    -\partial_l\Gamma^i{}_{jk}
    +\Gamma^m{}_{jl}\Gamma^i{}_{m k}
    -\Gamma^m{}_{jk}\Gamma^i{}_{m l}.
    \eea

    \def\ML{{\tilde L}}
    \def\MN{{\tilde N}}

    In the Euclidean formulation described above, the lapse function $N$
    is a real-valued function of time $t$. The standard formulation can be
    obtained by making $N$ imaginary, or equivalently by
    taking $N=i\MN$ with $\MN$ real.  Then the Euclidean Lagrangian $L$
    also becomes imaginary and so it is natural to describe the theory
    in terms of the real function $\ML=i\EL$, which is the Lagrangian
    for the standard formulation. It follows that the
    momenta and Hamiltonian in the standard formulation are related to their Euclidean
    counterparts (defined below) by identities of the form
    $\Pi_{standard}= i\Pi_{Euclidean}$ and $H_{standard}= i H_{Euclidean}$.
    Note also that in the standard formulation the Grassman variables $\psi^i$
    and $\overline\psi^i$ must be related by complex conjugation to ensure
    unitarity in the quantum theory. (This is unnecessary in the Euclidean
    version, where unitarity is not a requirement.)

    Having clarified its relationship with the standard formulation of the model,
    we henceforth consider only the Euclidean formulation described above.

    There are two different ways to define the (Euclidean) momenta conjugate
    to the variables $q^i$.
    If we regard the Lagrangian (\ref{eq:basic_Lagrangian}) as a function
    of the variables
    $q^i, \psi^i,\overline\psi_i$ and velocities $\dot q^i, \dot \psi^i,
    \dot{\overline\psi_i}$, then differentiating with respect to $\dot q^i$
    (while holding $\psi^i,\overline\psi_i, \dot\psi^i$ and
    $\dot{\overline\psi_i}$ fixed) gives
    \be
    p_i\equiv \left. {\partial \EL\over \partial \dot q^i}\right|_{\psi,\dot\psi}
    = N^{-1}g_{ij}\frac{dq^j}{dt}-\overline\chi\psi_i-\chi\overline\psi_i
    + \Gamma^j{}_{k i}\overline\psi_j\psi^k.
    \ee
    However, if we regard $\EL$ as a function of
    $q^i, \psi^i,\overline\psi_i, \dot q^i,$ and the {\it covariant}
    velocities $D\psi^i$ and $D\overline\psi_i$ and differentiate
    with respect to $\dot q^i$ (now holding $\psi^i,\overline\psi_i,
    D\psi^i$ and $D \overline\psi_i$ fixed) we instead obtain a set of
    covariantly defined momenta
    \be
    \Pi_i\equiv\left. {\partial \EL\over \partial q^i} \right|_{\psi,D\psi}
    = p_i - \Gamma^j{}_{k i}\overline\psi_j\psi^k.
    \ee
    These prove to be more useful for our purposes.

    From (\ref{eq:basic_Lagrangian}), we find that the
    momenta conjugate to $\chi$, $\overline\chi$ and $N$
    are not independent quantities,
    but are subject to the primary constraints
    \be
    \Pi_\chi \approx\Pi_{\bar\chi} \approx \Pi_N\approx 0. \label{eq:constraints1}
    \ee
    Since $\Pi_\chi$, $\Pi_{\bar\chi}$ and $\Pi_N$ vanish at all times,
    their time derivatives must also vanish.  After applying the equations
    of motion, this requirement gives rise to the secondary constraints
    \cite{Dirac}
    \be
    Q\approx \bar Q \approx {\cal H}\approx 0 \label{eq:secondary_constraints}
    \ee
    where we have defined the quantities\footnote{the ordering of the terms
    is immaterial here, but becomes more significant in the quantum theory.
    Different orderings may be used, and correspond to different choices
    of measure on the configuration space.}
    \be
    Q\equiv i\psi^i(\Pi_i+V_{,i}), \qquad
    \overline{Q}\equiv i(\Pi_i-V_{,i})\overline\psi^i
    \label{eq:susy_generators}
    \ee
    and
    \be
    {\cal H }\equiv -\frac{1}{2}g^{ij}(\Pi_i\Pi_j-V_{,i}V_{,j})
    +V_{;ij}\overline\psi^i\psi^j
    -\frac{1}{2}\overline R_{ijkl}
    \overline\psi^i\psi^j\overline\psi^k\psi^l.
    \label{eq:reparam_generator}
    \ee
    The constraint functions $Q,\bar Q$ and $\cal H$ are first class,
    and therefore generate of gauge symmetries.
    In fact, $Q$ and $\bar Q$ are the generators of supersymmetry transformations,
    while ${\cal H}$ is the generator of reparametrisations. The factors of $i$
    in the definitions (\ref{eq:susy_generators}) and the initial minus sign in
    the definition (\ref{eq:reparam_generator}) are included so that $Q$, $\bar Q$
    and ${\cal H}$ are identical to the constraints appearing
    in the conventional formulation of the model.

    The total Hamiltonian is then found to be just a linear combination of these
    constraints, and has the form
    \be
    H= -N ( {\cal H} + i\chi \overline Q + i\overline\chi Q)
    \label{eq:totalHamiltonian}
    \ee
    where $N$, $\chi$, and $\overline{\chi}$ can be regarded now as Lagrange
    multipliers. It vanishes when the constraints (\ref{eq:secondary_constraints})
    are imposed.

    There are also second-class constraints relating $\psi^i$ and
    $\bar\psi_i$ to their conjugate momenta. These constraints are
    interpreted as strong equalities and can be used to eliminate these
    particular momenta from the theory. Of course, the presence of
    second-class constraints requires us to employ Dirac brackets rather
    than the more familiar Poisson brackets.

    Following \cite{Casalbuoni}, the elementary Dirac brackets are found to
    be simply
    \be
    \{q^i,q^j\}=0,\qquad  \{q^i,\Pi_j\} = \delta^i{}_j , \qquad
    \{\Pi_i,\Pi_j\}= \overline\psi_k\psi^l R^k{}_{lij}
    \ee
    \be
    \{\psi^i,\psi^j\}=0, \qquad
    \{\overline\psi_i,\overline\psi_j\}=0, \qquad
    \{\psi^i,\overline\psi_j\} = \delta^i{}_j
    \ee
    \be
    \{q^i,\psi^j\} =0, \quad \{q^i,\overline\psi_j\}=0,\quad
    \{\Pi_i,\psi^j\}= \Gamma^j{}_{ki}\psi^k, \quad
    \{\Pi_i,\overline\psi_j\}=-\Gamma^k{}_{ji}\overline\psi_k
    \ee
    from which we obtain
    \be
    \qquad \{Q,{\cal H}\} = 0,
    \qquad \{\overline Q,{\cal H}\}=0.
    \ee
    and
    \be
    \{Q,\overline{Q}\} = 2{\cal H}.\label{eq:susy_bracket}
    \ee
    These equations are the hallmark of supersymmetry.

    Defining the fermion number $F\equiv \overline\psi_i \psi^i$, we also have
    \be
    \{F,\overline{\psi}_j\}=\overline{\psi}_j \qquad  \{ F,\psi_j\}=-\psi_j
    \ee
    while
    \be
    \{F,\overline Q\} = \overline Q  \qquad  \{F,Q\}= -Q, \qquad \{F,{\cal H}\}=0
    \ee
    showing that supersymmetry transformations do not mix fermion number,
    and that fermion number is unaffected by reparametrisations. Moreover,
    fermion number is a constant of the motion, since
    \be
    \{F,H\}\approx 0.
    \ee

    We conclude this section by remarking that in the case of rigid
    supersymmetry, $\overline \chi(t), \chi(t)$ and $N(t)$ are specified
    functions rather than Lagrange multipliers. Hence the supersymmetry
    and reparametrisation generators $Q,\overline Q$ and ${\cal H}$
    are not constrained to vanish, and nor is the total
    Hamiltonian (\ref{eq:totalHamiltonian}).

    \section{Boundary corrections to the action}\label{BoundaryTerms}

    Supersymmetry is always broken when boundaries are present, since
    the supersymmetric variation of the Lagrangian is a total divergence
    which yields a boundary term when integrated.
    In this section we will see that there is a boundary correction to
    the action that restores invariance under a sub-algebra of
    supersymmetry transformations \cite{BoundaryCorrections}.

    The classical trajectories are invariant under the following
    infinitesimal supersymmetry transformations:
    \bea
    \delta q^i &=&\overline{\epsilon}\psi^i+ \epsilon{\overline\psi}^i
    \label{eq:trans1}\\
    \delta\psi^i &=&
    -\epsilon g^{ij} ( \Pi_i -V_{,i}) - \Gamma^i{}_{jk} \psi^j \delta q^k\\
    \delta\overline\psi_i &=&-\overline\epsilon (\Pi_i + V_{,i})
    + \Gamma^j{}_{ik} \overline\psi_j \delta q^k\\
    \delta \chi &=&N^{-1}\dot{\epsilon}+2\epsilon\overline{\chi}\chi\\
    \delta \overline{\chi}
    &=&N^{-1}\dot{\overline{\epsilon}}+2\overline{\epsilon}\chi\overline{\chi}\\
    \delta N &=&-2(\overline{\epsilon}\chi+\epsilon\overline{\chi})\label{eq:trans7}
    \eea
    where $\epsilon$, $\overline{\epsilon}$ are anticommuting Grassmann
    variables. The variation of the action is given by the boundary term
    \bea
    \delta S={1\over 2}\Bigl[
    \epsilon\overline{\psi}_i(N^{-1}\dot q^i +g^{ij}V_{,j}
    -\overline{\chi}\psi^i)
    +\overline{\epsilon}\psi^i(N^{-1}g_{ij}\dot q^i -V_{,i}-\chi
    \overline\psi_i)\Bigr]_{t_1}^{t_{2}}.
    \label{eq:bdry_terms}
    \eea
    While it is customary to disregard boundary terms, we cannot afford to do
    so here. The supersymmetry transformation is broken by the
    boundary terms (\ref{eq:bdry_terms}), preventing us from
    using Nicolai's theorem. In order to be able to construct a Nicolai map,
    we must find a way to restore the invariance of the action under
    a sub-algebras of the supersymmetry generators \cite{Luckock:2}.

    Exact invariance of the action under a sub-algebra of
    supersymmetry generators can be restored if the Lagrangian
    is augmented or diminished by the total derivative
    \be
    L_B=\frac{d}{dt}(V+\frac{1}{2}\overline\psi_i\psi^i).
    \ee
    Indeed, under the infinitesimal supersymmetry transformations
    (\ref{eq:trans1}-\ref{eq:trans7}) the
    variation of the integral
    $$I \equiv \int_{t_1}^{t_2} L_B\, dt\,  = \,
    \left[V+\frac{1}{2}\overline{\psi}_i\psi^i \right]_{t_1}^{t_2} $$
    is found to be
    \be
    \delta I={1\over 2}\Bigl[
    \epsilon\overline{\psi}_i(N^{-1}\dot q^i+g^{ij}V_{,j}
    -\overline{\chi}\psi^i)
    -\overline{\epsilon}\psi^i(N^{-1}g_{ij}\dot q^i-V_{,i}-\chi
    \overline{\psi}_i)\Bigr]_{t_{1}}^{t_{2}}.
    \ee
    Consequently, the supersymmetric variations of the modified actions
    \be
    S_\pm = S\pm I = \int_{t_1}^{t_2} (\EL\pm L_B) dt \label{eq:mod_actions}
    \ee
    are
    \be
    \delta S_+=\Bigl[
    \epsilon\overline{\psi}_i(N^{-1}\dot q^i+g^{ij}V_{,j}
    -\overline{\chi}\psi^i)\Bigr]_{t_1}^{t_2}
    \ee
    and
    \be
    \delta S_-=\Bigl[\overline{\epsilon}\psi^i(N^{-1}g_{ij}
    \dot q^i-V_{,i}-\chi \overline{\psi}_i)\Bigr]_{t_{1}}^{t_{2}}
    \ee
    If we specify that $\epsilon(t_1)=\epsilon(t_2)=0$, then $\delta S_+$
    will vanish exactly. In other words, the modified action $S_+$ is exactly
    invariant under the subalgebra of infinitesimal supersymmetry transformations
    obtained by imposing Dirichlet boundary conditions on $\epsilon$. We will
    refer to this as the {\it left-handed subalgebra}.

    Similarly,  the modified action $S_-$ is exactly
    invariant under the subalgebra of infinitesimal supersymmetry
    transformations obtained by imposing Dirichlet boundary conditions on
    $\overline\epsilon$. We will refer
    to this as the {\it right-handed subalgebra}.

    Because the modified actions differ from the original only by boundary
    terms, the classical equations of motion are unaffected. (In fact, by adding
    a boundary correction to the action, we have
    really just performed a canonical transformation.)  However, the
    different versions of the action give rise to different expressions for
    the momenta. If we use the modified action $S_+=\int (\EL+L_B)$
    (which is invariant under the left-handed subalgebra of supersymmetry
    generators), then the covariant momentum conjugate to $q^i$ is
    $$ \Pi_i^+ \equiv \left. {\partial (\EL+L_B)\over \partial \dot q^i}
    \right|_{\psi,\bar\psi,D\psi,D\bar\psi}
    =\Pi_i+ V_{,i}$$
    and we can write the supersymmetry generators as
    \be
    Q=i\psi^i\Pi_i^+ , \ \  \ \overline Q= i(\Pi_i^+ - 2V_{,i})\overline \psi^i
    .\label{eq:plusrep}
    \ee
    On the other hand, if we use the modified action $S_-=\int (\EL-L_B)$ (which is
    invariant under the
    right-handed subalgebra of supersymmetry generators), then the
    covariant momentum conjugate to $q^i$ is
    $$ \Pi_i^- \equiv \left. {\partial (\EL-L_B)\over \partial \dot
    q^i}\right|_{\psi,\bar\psi,D\psi,D\bar\psi}
    =\Pi_i- V_{,i}$$
    and we can write the supersymmetry generators as
    \be
    Q=i\psi^i(\Pi_i^- +2V_{,i}) , \ \  \ \overline Q= i\Pi_i^-\overline \psi^i
    .\label{eq:minusrep}
    \ee

    Quantisation of the theory then involves representing the canonical
    variables as operators and interpreting $t$. However before doing this
    let us briefly consider the special case of rigid supersymmetry.

    \section{Rigid Supersymmetry and Nicolai Maps}\label{FokkerPlanck}
    The local symmetry described above reduces to rigid supersymmetry if
    one fixes the values of the (non-dynamical) Lagrange multipliers as
    (for example)
    \be N=1, \qquad \chi=0, \qquad \bar\chi=0.\ee
    In order that these values are preserved under supersymmetry
    transformations, the transformation parameters must be constant:
    \be \dot \epsilon = 0 ,\ \ \ \ \ \dot {\bar\epsilon} = 0 .\ee

    Because $N,\chi,\bar\chi$ have fixed values and are not allowed to
    vary, they no longer act as Lagrange multipliers enforcing first-class
    constraints. Hence, when the supersymmetry is rigid, the quantities
    $Q,\overline Q$ and $\cal H$ defined in equations
    (\ref{eq:susy_generators},\ref{eq:reparam_generator}) are no longer
    required to vanish. As a consequence, the Hamiltonian may also be
    non-vanishing.

    The two invariant forms of the action now reduce to
    \be
    S_+=\int_{t_1}^{t_2}
    \Bigl\{ \textstyle{1\over 2} g_{ij}( \dot q^i + V^{,i})(\dot q^j + V^{,j}) +
    \overline\psi_iD\psi^i
    +V_{;ij} \overline\psi^i\psi^j -\textstyle{1\over 2} R_{ijkl}
    \overline\psi^i\psi^j\overline\psi^k\psi^l \Bigr\} dt
    \ee
    and
    \be
    S_-= \int_{t_1}^{t_2}
    \Bigl\{ \textstyle{1\over 2} g_{ij}( \dot q^i - V^{,i})(\dot q^j -V^{,j})
    -(D\overline\psi_i)\psi^i
    +V_{;ij} \overline\psi^i\psi^j -\textstyle{1\over 2} R_{ijkl}
    \overline\psi^i\psi^j\overline\psi^k\psi^l \Bigr\} dt
    \ee
    According to Nicolai's theorem \cite{Nicolai}, the invariance of each of
    these actions under a subalgebra of supersymmetry generators ensures that
    the theory can be transformed into a free bosonic theory by integrating out
    the fermions. The transformation will generally have the form of a
    first-order differential equation, and so will only be invertible if
    initial conditions are imposed on the bosonic variables $q^i(t)$ in the
    interacting theory. However, for Nicolai's theorem to work, any initial
    conditions must be invariant under the same supersymmetry subalgebra as
    the action \cite{Luckock:2}. For the action $S_+$, appropriate initial
    conditions are
    \be
    q^i(t_1)= q^i_1,\ \ \ \ \ \ \psi^i(t_1) =0  \label{eq:QinvIC}
    \ee
    since these (like $S_+$ itself) are invariant under the left-handed
    subalgebra. For $S_-$ one must use instead the initial conditions
    \be
    q^i(t_1)= q^i_1,\ \ \ \ \ \ \ \bar\psi_i(t_1) =0.\label{eq:QbarinvIC}
    \ee

    Integrating out the fermions (subject to the appropriate initial conditions),
    one finds that the weight of a particular path $q(t)$ in the ensemble of
    possible paths is given by \cite{Graham:1985}
    \bea
    {\cal P}_\pm[q(t)]&=& \int [\psi,\bar\psi] \exp \left( -{1\over\hbar}
    S_\pm[q,\psi,\bar\psi] \right) \cr\cr
    &=& J_\pm[q]\exp \left\{ - {1\over \hbar} \int_{t_1}^{t_2}
    \textstyle{1\over 2}
    g_{ij}( \dot q^i\pm V^{,i})(\dot q^j \pm V^{,j}) \right\}
    \eea
    where
    \be
    J_\pm[q]\equiv \exp \left\{\pm {1\over 2\hbar}
    \int_{t_1}^{t_2} [g^{ij} V_{;ij} -\textstyle{1\over 4} R ]\right\}  dt
    \ee
    and $R(\q)$ denotes the curvature scalar of the configuration
    space at the point $\q(t)$. In fact the functional
    $J_\pm[q]$ turns out to be the Jacobian for the transformation
    $\xi^a(t)\mapsto q^i(t)$ defined by the differential equation
    \be
    {\Delta q^i\over dt} = \mp g^{ij}V_{,i} \ + e^i{}_a\cdot \xi^a
    \label{eq:Langevin}
    \ee
    where $e^i{}_a$ is a vielbein field on the configuration space with
    the property  that
    \be
    g_{ij} e^i{}_a e^j{}_b =\delta_{ab}
    \ee
    and the differentials $\Delta q^i =dq^i+ g^{jk} \Gamma^i{}_{jk}dt$ and
    $e^i{}_a\cdot \xi^a dt$ are defined so that they transform covariantly in
    the It\^o calculus. It follows by a change of variable that the weight
    for a given history $\xi^a(t)$ of the new variable is
    \be
    {\cal P}[\xi(t)] = \exp\left\{
    -{1\over\hbar} \int_{t_1}^{t_2} {1\over 2} \delta_{ab} \xi^a\xi^b dt \right\}
    \ee
    and so $\xi^a(t)$ can be interpreted as a white noise process with
    auto-correlation
    \be
    \langle \xi^a(t)\xi^b(t') \rangle = {\hbar\over 2}  \delta(t-t')
    \ee
    that drives the motion of the particle in configuration space via the Langevin
    equation (\ref{eq:Langevin}). The probability density function $f_\pm(t,\q)$
    for the particle's position $\q$ at time $t$ will then evolve according
    to the associated Fokker-Planck equation
    \be
    {\partial f_\pm\over\partial t} = {\partial\over \partial q^i}
    \left[  g^{1/2} g^{ij}\left( {\hbar\over 2} {\partial f_\pm\over \partial q^j}
    \, \pm
    V_{,j} f_\pm \right) \right] \label{eq:FokkerPlanck}
    \ee
    where $g$ denotes the determinant of the matrix of components $g_{ij}$ of
    the configuration space metric.

    In fact this result can be obtained directly from canonical quantisation.
    The evolution of the quantum state vector $|\Psi(t)\rangle$ is governed by
    the Schr\"odinger equation, which in this case can be written
    \be
    {d \over dt} | \Psi(t)\rangle = {1\over\hbar} H  | \Psi(t)\rangle
    = - {1\over 2\hbar^2} Q \overline Q | \Psi(t)\rangle \ -
    {1\over 2\hbar^2} \overline Q Q | \Psi(t)\rangle \label{eq:Schrodinger}
    \ee
    on account of the operator identity
    $Q\overline Q+\overline QQ= -2\hbar H$
    that follows from the form of the Dirac brackets and our choices of values
    for the Lagrange multipliers.

    As remarked above, Nicolai's theorem is applicable only if the action
    and initial conditions are invariant under one of the supersymmetry
    subalgebras. Requiring invariance under the left-handed subalgebra
    generated by $Q$ means that the initial state $|\psi(t_1)\rangle$
    must be annihilated by the operator $Q$; and since this operator
    commutes with the Hamiltonian it follows that $Q |\psi(t)\rangle=0$
    for all $t$ and hence
    \be {d \over dt} | \Psi(t)\rangle =-{1\over 2\hbar^2}
    Q\overline Q |\Psi(t)\rangle
    =  {1\over 2\hbar^2} \psi^i \Pi_i^+ (\Pi_j^+ -2V_{,j})\overline \psi^j
    \, |\Psi(t)\rangle  \label{eq:Schro1}
    \ee
    where we have used (\ref{eq:plusrep}) to write $Q$ and $\overline Q$ in terms
    of the covariant momenta $\Pi_i^+$ associated with the $Q$-invariant
    action $S_+$. In fact these momenta are naturally represented by
    $-\hbar \nabla_i$, where $\nabla_i$ denotes the covariant derivative
    with respect to $q^i$.

    In general the wave function will have a number of components,
    corresponding to solutions with different fermion numbers. Each
    such component is naturally represented as a $p$-form on the
    configuration space, where $n-p$ is the fermion number
    \cite{Witten,AlvarezGaume}.
    The fermion annihilation and creation operators $\psi^i$ and
    $\overline \psi_i$ then act on a form $\omega$ according to
    \be
    \psi^i \omega = dq^i\wedge \omega, \ \ \ \ \ \ \  \
    \overline \psi_j \omega  = \hbar i_j\omega
    \ee
    where $i_j\omega $ denotes the contraction of the form $\omega$ with the
    vector field $\partial/\partial q^j$. With the momenta $\Pi^-_i$
    represented in the manner described above, it follows that
    \be
    \psi^i \Pi_i^+ \omega =-\hbar d\omega,\ \ \ \ \
    \Pi_i^+ \overline\psi^i \omega =\hbar^2 \delta \omega
    \ee
    where $d$ is the exterior derivative and $\delta$ is it adjoint,
    the coderivative. So if the state $|\Psi(t)\rangle$ is represented by the form
    $\omega$, then (\ref{eq:Schro1}) implies
    \be
    {\partial\omega \over\partial t} =-\textstyle{1\over 2}
    d (\hbar  \delta\omega - 2 i_V\omega)
    \label{eq:FP1}
    \ee
    where $i_V$ denotes the contraction with the vector field $g^{ij} V_{,j}
    \partial /\partial q^i$. The $Q$-invariant initial condition
    (\ref{eq:QinvIC}) implies $\psi^i|\Psi(t_1)\rangle=0$ and
    hence $F|\Psi(t)\rangle=0$ for all $t$
    (since the fermion number $F=\overline\psi_i\psi^i$ is conserved).
    Consequently, $\omega$ will be an $n$-form, and at each point in
    the configuration
    space must be proportional to the elementary $n$-form:
    \be \omega= f_+(\q)  dq^1\wedge dq^2\wedge \ldots \wedge dq^n .\ee
    It is then easily verified that (\ref{eq:FP1}) agrees precisely with
    equation (\ref{eq:FokkerPlanck}) for $f_+(\q)$.

    Alternatively, if one starts with the $\overline Q$-invariant
    action $S_-$ and initial conditions (\ref{eq:QbarinvIC}) then
    the Schr\"odinger equation (\ref{eq:Schrodinger}) reduces to
    \be {d \over dt} | \Psi(t)\rangle =- {1\over 2\hbar^2}
    \overline Q Q |\Psi(t)\rangle
    =  {1\over 2\hbar^2} \Pi_i^- \overline\psi^i \psi^j (\Pi_j^- +2V_{,j})
    \, |\Psi(t)\rangle  \label{eq:Schro2}
    \ee
    and so if the state is represented by a differential form $\omega$ then
    \be
    {\partial \omega\over \partial t} =
    - \textstyle{1\over 2} \delta (\hbar d\omega -2dV\wedge \omega).
    \label{eq:FP2}
    \ee
    (Note that in this representation it is the momenta $\Pi_i^-$, rather than
    that are $\Pi_i^+$, that are represented by the operator $-\hbar \nabla_i$;
    hence $\psi^i\Pi_i^-\omega = -\hbar d\omega$ and
    $\Pi_i^-\overline\psi^i\omega=\hbar^2 \delta\omega$.)
    In this case, the initial conditions $\overline\psi^i |
    \Psi(t_1)\rangle=0$ imply that $\omega$ is a 0-form; say
    $\omega = f_-(\q)$ and so this equation reduces to (\ref{eq:FokkerPlanck})
    with the minus sign chosen. The evolution equation (\ref{eq:FP1}) is
    then seen to be equivalent to (\ref{eq:FokkerPlanck}) for $f_-(\q)$.

    Which of the two representations is more appropriate depends on the form
    of the configuration space potential $V(\q)$. For definiteness, we
    suppose that the potential grows as $|\q|$ increases such that
    $\lim_{|\q|\to\infty} V(\q)/\ln|\q| >2\hbar n$. In this case, the
    Fokker-Planck equation (\ref{eq:FokkerPlanck}) admits a static normalisable solution
    \be
    f_+(\q) = A \exp \left( -{V(\q) \over 2\hbar} \right) \label{eq:fplus}
    \ee
    that represents the limiting $t\to\infty$ probability density function of
    a Brownian particle driven by the Langevin equation
    \be
    {\Delta q^i\over dt} = - g^{ij}V_{,i} \ + e^i{}_a\cdot \xi^a.
    \ee

    \section{Quantisation with Local Supersymmetry}\label{localSUSY}
    Our goal in this section is to quantise the locally supersymmetric
    model, and to interpret the resulting theory. However we start by
    reviewing the method described above for rigid supersymmetry (albeit
    in slightly different notation).

    If the model has only {\it rigid} supersymmetry,
    quantisation means finding solutions of the Schr\"odinger equation
    \be
    {d\over dt} |\Psi(t)\rangle = {1\over \hbar} H\, |\Psi(t)\rangle\label{eq:Schro}
    \ee
    with $H=-N({\cal H} + i \chi\overline Q+ i \overline\chi Q)$
    and fixed values of $\bar\chi$, $\chi$ and $N$.
    For the Euclidean theory we choose $N=1$ and $\bar\chi=\chi=0$ ,
    and so the general solution has the form
    \be
    |\Psi(t)\rangle = e^{{-t {\cal H}/\hbar}} |\Psi_0\rangle\label{eq:gensol}
    \ee
    with $|\Psi_0\rangle$ an arbitrary initial state.

    Assuming that $V(\q)$ grows as $|\q|$ increases,
    we choose the initial state $|\Psi_0\rangle$ to be invariant under
    left-handed supersymmetry; $Q |\Psi_0\rangle=0$.
    Because $[{\cal H} , Q]=0$, it follows that
    \be Q |\Psi(t)\rangle=0 \ \ \ \ \forall t\ge 0.
    \label{eq:Qinv_td}
    \ee
    As discussed in the last section, in this case
    (\ref{eq:Schro}) can then be represented as a Fokker-Planck
    equation describing the evolution of a conserved probability
    distribution.

    The static probability distribution can be obtained by taking the
    $t\to\infty$ limit. (This limit must exist, due to the assumed
    form of the potential $V(\q)$.) In the current notation, this
    is represented by the state
    $$ |\Psi_\infty \rangle = \lim_{t\to\infty} | \Psi(t) \rangle.$$
    The time-independence of this state implies that it has zero energy;
    $$ {\cal H} |\Psi_\infty \rangle = 0.\label{eq:static}$$
    (Note that in this respect the Euclidean theory considered
    here differs from the conventional theory, which admits a
    ground state with a positive energy and hence an oscillating
    phase. However, because the Euclidean evolution operator
    $e^{{-t {\cal H}/\hbar}}$ is hermitian rather than
    anti-hermitian, only a zero-energy state can have
    a constant norm.)

    Thanks to (\ref{eq:Qinv_td}), this limiting state is also
    $Q$-invariant:
    \be
    Q |\Psi_\infty\rangle=0.
    \ee
    Moreover, since $0=\langle \Psi_\infty |2\hbar {\cal H} | \Psi_\infty\rangle
    = \langle \Psi_\infty |Q\overline Q | \Psi_\infty\rangle +
    \langle \Psi_\infty |\overline Q Q | \Psi_\infty\rangle
    =0 + ||\,\overline Q | \Psi_\infty\rangle ||^2$,
    this state must also be $\overline Q$-invariant:
    \be
    \overline Q |\Psi_\infty\rangle=0.
    \ee
    This is readily verified in the Fokker-Planck
    representation considered earlier; one has only to show that the function $f_+(\q)$
    given by (\ref{eq:fplus}) is annihilated by the operator $\overline Q$
    defined in (\ref{eq:plusrep}).

    The state $|\Psi_\infty\rangle$ is thus annihilated by all the
    first-class constraints of the {\it locally} supersymmetric theory
    and thus represents an acceptable quantum state for this theory also.

    Let us summarise the argument so far. Thanks to Nicolai's theorem
    the supersymmetry of the model ensures that there exists a representation
    of the quantum theory in which the Euclidean Schr\"odinger equation
    takes the form of a Fokker-Planck equation governing the evolution of
    a conserved probability density function. By taking the (Euclidean)
    long-time limit one obtains a time-independent state satisfying all the
    constraints of the locally supersymmetric theory and represented by
    a static probability density function. This procedure is readily
    adapted (at least in principle) to quite general theories with local
    supersymmetry, and is naturally regarded as a type of stochastic
    quantisation.

    Apart from giving a procedure for finding quantum state, this approach
    also provides a natural probabilistic interpretation of this state
    space without reference to any particular inner product. Indeed, by
    reviewing the derivation of the probability density function (\ref{eq:fplus}),
    the reader can confirm that the inner product has not been used at all.
    This is potentially of importance for theories such as supergravity,
    in which the choice of
    the inner product is itself problematic. By using the method outlined
    above, one can (in principle) obtain not only the quantum state that
    satisfying all the constraints of the theory but also the
    corresponding probability density function on the configuration space.

    \section{Discussion and conclusion}

    In this paper, stochastic quantisation of a locally supersymmetric model
    is performed without introducing an extra fictitious time variable as is
    normally done in stochastic quantisation.

     In the context of supersymmetric models, Nicolai maps
    exist for certain components of the wave function. This suggests an
    interpretation of the wave
    function as a probability density whose integrals is conserved in
    Euclidean time. In the quantum theory, the
    bosonic components of the wave function, satisfy a type of
    Fokker-Planck equation and thus can be interpreted stochastically.
    However, since our supersymmetric theory is local one can by-pass
    the difficulties of explicitly constructing Nicolai maps by finding
    the associated Fokker-Planck equation directly via canonical quantisation
    with a stochastic interpretation.

    In conventional quantisation of a classical theory one has to construct
    the Hilbert space and define an inner product in the Hilbert space.
    The probability then is the squared modulus of the wave function.
    However, in the stochastic quantisation of locally supersymmetric
    theory, the solution of the Fokker-Planck equation is a probability
    density function. Thus, there is a ready-made probabilistic interpretation
    of the quantum theory without any reference to the concept of inner product
    as one see in the usual Hilbert space formulation of quantum theory.
    This should be a big advantage in quantum gravity and cosmology,
    where the choice of satisfactory inner product has long been viewed
    as one of the most fundamental problems.

    \end{document}